\documentclass[12pt,a4paper]{article}
\usepackage{graphicx}% Include figure files
\usepackage{dcolumn}% Align table columns on decimal point
\usepackage{bm}% bold math
\usepackage{color}
\def\bea{\begin{eqnarray}}
\def\eea{\end{eqnarray}}

\newcommand{\nn}{\nonumber}

\def\beq{\begin{equation}}
\def\eeq{\end{equation}}

\usepackage{amssymb}
\usepackage{amsmath}

\begin{document}

%\begin{flushright}
%hep-th/9903135\\
%NUP-A-98-8 \\
%\end{flushright}
%\hspace{\fill}

\vspace{14mm}
\begin{center}
        \Large\bf A Vanishingly Small  Vector Mass from Anisotropy of Higher Dimensional Spacetime
\end{center}
\hspace{\fill}
\begin{center}
        {\large
                Taegyu Kim and Phillial Oh\\[3mm]
                
                {\it Department of Physics and Institute of Basic Science,\\
                        Sungkyunkwan University, Suwon 440-746, Korea\\[2mm]

                        {\tt taegyukim@skku.edu,~ploh@skku.edu}}}
        
\end{center}

\hspace{\fill}
\vspace{5mm}
\begin{flushleft}
        \bf{Abstract}
\end{flushleft}

We consider five-dimensional massive vector-gravity theory which is based on the foliation preserving diffeomorphism and anisotropic conformal invariance. It does not have an intrinsic scale and the only relevant parameter is the anisotropic factor $z$ which characterizes the degree of anisotropy between the four-dimensional spacetime and the extra dimension. We assume that physical scale $M_*$ emerges as a consequence of spontaneous conformal symmetry breaking of vacuum solution. It is demonstrated that a very small mass for the vector particle compared to $M_*$ can be achieved with a relatively mild adjustment of the parameter $z$. At the same time, it is also observed that the motion along the extra dimension can be highly suppressed and the five-dimensional theory can be effectively reduced to four-dimensional spacetime.

\vspace{4mm}

%\noindent
%PACS numbers: 

\newpage
%
        %Contents
        \section{Introduction}
        The research on the non-vanishing photon mass has a long history \cite{Tu:2005ge}. It is well known that Maxwell theory with massless photon can be extended to a gauge-invariant massive Stueckelberg theory by introducing a scalar field that compensates the gauge transformation of the vector field. It preserves the unitarity and renormalizability of the massless theory and describes electrodynamics of massive vector field \cite{Ruegg:2003ps}. In a particular gauge where the scalar field is set to zero, the theory reduces to the familiar Proca theory \cite{Accioly:2010zzb}. The deviation of photon mass from zero, if any, must be extremely small because Maxwell theory has been tested to  high experimental accuracy. The experimental constraints on the photon mass have considerably increased over the past several decades, putting stringent upper bounds on its mass. Recently, the cosmological implication of massive photon was investigated in connection with the dark energy and it was argued that current acceleration is provided by the non-vanishing photon mass $m$ governed by the relation $\Lambda\sim m^2$ \cite{Kouwn:2015cdw}.
        
        Up to now, there does not exist a method of generating an extreme small mass of vector particles in general, and the mass has to be put in by hand if you start in four-dimensional spacetime. This requires  extreme fine tuning and from a theoretical point of view, it would be aesthetically more appealing if this could be avoided. One way of bypassing this problem is to consider higher-dimensional theory in which gravity is coupled with five-dimensional Maxwell-Stueckelberg theory. However, there is a caveat here. It turns out that usual higher dimensional theories in which the spacetime is isotropic can not provide the method which suppresses the photon mass to a high degree. The fine-tuning problem still remains.  It can be shown that a possible remedy necessitates a higher dimensional theory in which the four-dimensional spacetime and extra dimensions are not treated on an equal footing \cite{Moon:2017rox}. On top of it, the theory requires two compatible symmetries of foliation preserving diffeomorphism (FPD) and anisotropic conformal transformation (ACT). The FPD is implemented in the ADM decomposition of higher dimensional metric by requiring the foliation preserving diffeomorphism invariance adapted to the extra dimensions, thus keeping the general covariance only for the four dimensional spacetime. Conformal invariance can be incorporated with an extra (Weyl) scalar field and a real parameter $z$ which describes the degree of anisotropy of conformal transformations between the four-dimensional spacetime and extra dimensional metrics. A cosmological test of $z$ was given yielding an allowed range of the parameter \cite{Kouwn:2017qet}.   
               
                In this paper, we construct five-dimensional Stueckelberg-like vector theory in an anisotropic spacetime background and couple it with the gravity theory which has the symmetry of the aforementioned FPD and ACT. The essential point of the construction in the gravity sector is a parameter $z$ which describes the degree of anisotropy of conformal transformation between the four-dimensional spacetime and extra dimension. This construction can be extended to the higher dimensional vector theory straightforwardly and the coupled action describes five-dimensional vector-gravity theory which respects both FPD and ACT. The theory is so constructed as not to contain any intrinsic scale. Then, we look for a flat Minkowski vacuum solution with symmetry breaking scale $M_*$ and discuss possible emergence of vector mass which is vastly suppressed compared to  $M_*$. It turns out that this is possible for a suitable adjustment of the parameter $z$. The degree of fine-tuning is  alleviated to a great extent. Another interesting consequence of the analysis is that the propagation of the massive vector field along the fifth dimension is highly suppressed and makes the extra dimension obsolete as far as dynamics is concerned. Thus it could reduce the higher dimensional spacetime to effectively four spacetime for a certain range of the parameter $z.$  
                
                This paper is organized as follows. In section 2, we construct five-dimensional   Stueckelberg-like vector-gravity theory with anisotropic conformal invariance.  In Sec. 3, we search for the Minkowski vacuum solution and consider the effective action of the massive vector field sector. A gauge fixing term is introduced and propagators are presented. Sec. 4 contains conclusions and discussions. 
        
        \section{Model}
        We start with a formulation of 5D anisotropic conformal gravity.
        The  first part of this section is mostly redrawn from Ref. \cite{Moon:2017rox, Kouwn:2017qet}
        to make the paper self-contained. Let us first consider the
        Arnowitt-Deser-Misner (ADM) decomposition of five-dimensional metric:
        \begin{align}
        ds^2 = g_{\mu\nu}( dx^\mu +N^\mu dy)( dx^\nu +N^\nu dy) + N^2 dy^2. \label{model}
        \end{align}
        Then, the five-dimensional Einstein-Hilbert action without cosmological constant is expressed as
        \begin{align}
        S^{(5)}_{\rm{EH}}=\int \,
        d^4x dy N\sqrt{-g}~M_*^3\left[R-\{K_{\mu\nu}K^{\mu\nu} -
        K^2\}\right],
        \label{5dR}
        \end{align}
        where $M_*$ is the five dimensional gravitational constant, $R$ is the spacetime curvature and $K_{\mu\nu}$ is the extrinsic curvature tensor,
        $K_{\mu\nu}=(\partial_y g_{\mu\nu} - \nabla_{\mu} N_{\nu} -
        \nabla_{\nu} N_{\mu})/(2N)$.
        The above action (\ref{5dR}) can be extended anisotropically
        by breaking the five dimensional general covariance down to its foliation preserving
        diffeomorphism symmetry given by
        \begin{align}
        x^{\mu}\to x^{\prime\mu}&\equiv x^{\prime\mu}(x,y),
        ~~~y\to y^{\prime}\equiv
        y^{\prime }(y),\label{FPD}\\
        g^{'}_{\mu\nu}(x',y')&=\left(\frac{\partial x^\rho}{\partial
                x^{'\mu}}\right)\left(\frac{\partial
                x^\sigma}{\partial x^{'\nu}}\right)g_{\rho\sigma}(x,y),\label{trans1}\\
        {N'}_{}^{\mu}(x',y')&=\Big(\frac{\partial y}{\partial
                y^{'}}\Big)\Big[\frac{\partial x^{'\mu}}{\partial
                x^{\nu}}N_{}^{\nu}(x,y)-\frac{\partial x'^{\mu}}{\partial
                y^{}}\Big],\label{trans2}\\
        N^{\prime}(x',y')&=\left(\frac{\partial y}{\partial y{'}}\right)N(x,y)
        , \label{trans3}
        \end{align}
        and non-uniform conformal transformations
        \begin{eqnarray}
        g_{\mu\nu}\rightarrow e^{2\omega(x,y)}g_{\mu\nu},~~
        N\rightarrow e^{z \omega(x,y)}N,~~ N^{\mu}\rightarrow N^{\mu},
        ~~\phi\rightarrow
        e^{-\frac{z+2}{2}\omega}\phi,\label{confotrans11}
        \end{eqnarray}
        where a Weyl scalar field $\phi$ to compensate the conformal transformation
        of the metric is introduced.
        In the above equation \eqref{confotrans11}, a factor $z$ is introduced in the transformation of $N(=g_{55})$, 
        which characterizes the anisotropy of spacetime and
        extra dimension.
        The anisotropic Weyl action invariant under Eqs. \eqref{FPD}-\eqref{confotrans11} for an arbitrary $z$ can be written as
        %\footnote{
        %We could consider more general action with foliation preserving diffeomorphism %invariance where the foliation is adapted to extra dimension \cite{}. But %the action (\ref{conformalR}) exhibits all the essential features as far %as cosmology is concerned.}
        \begin{eqnarray}
        &&\hspace*{-2em}S^{(5)}_{\text{CG}}=\int d^4x dy
        \sqrt{-g}N\Bigg[\phi^{2}\left(R
        -\frac{12}{z+2}\frac{\nabla_{\mu}\nabla^{\mu}\phi}{\phi}
        +\frac{12 z}{(z+2)^2}\frac{\nabla_{\mu}\phi\nabla^{\mu}\phi}{\phi^2}
        \right)\nonumber\\
        &&\hspace*{7em}~
        -\beta_1\phi^{-\frac{2(z-4)}{z+2}}\left\{B_{\mu\nu}B^{\mu\nu} -
        \lambda B^2\right\}+\beta_2 \phi^{2}C_{\mu}C^{\mu}\Bigg].\label{conformalR}
        \end{eqnarray}
        where $\beta_1, \beta_2, \lambda$ are some constants,
         $B_{\mu\nu}$ and $C_\mu$ are given by
        \begin{align}
        B_{\mu\nu}&=K_{\mu\nu}+\frac{2}{(z+2)N\phi}g_{\mu\nu}(\partial_y
        \phi-\nabla_{\rho}\phi N^{\rho})\,,\quad
        B\equiv g^{\mu\nu}B_{\mu\nu},\\
        C_{\mu}&=\frac{\partial_{\mu}N}{N}
        +\frac{2 z}{z+2}\frac{\partial_{\mu}\phi}{\phi}.
        \end{align}
        
        Several comments are in order. Note that the above action (\ref{conformalR}) is built so as not to contain any intrinsic scale in accordance with conformal symmetry. The scale $M_*$ of (\ref{5dR})  is supposed to emerge as consequence of spontaneous conformal symmetry breaking. We consider only the case $z \neq -2$, because $\phi$ is not effected
        under the conformal transformation in \eqref{confotrans11} in this case.
        It can be actually shown that for  $z = -2$, an anisotropic scale invariant gravity
        theory can be constructed without the need of the field $\phi$. In this work, we are interested in Minkowski vacuum solution and we did not include the potential term in the above action (\ref{conformalR}). The isotropic case with $\beta_1=\lambda=z=1,$ and  $\beta_2=0$ leads to five dimensional Weyl gravity with a zero potential for the field $\phi$ \cite{Moon:2017rox, Kouwn:2017qet}. In the anisotropic case, the action (\ref{conformalR})
        is, in general, plagued with perturbative ghost instability coming from breaking of the full general covariance of 5D. However, it can be shown that this problem can be cured by constraining  the constants  $\beta_1$ and $\beta_2$, especially with $0<\beta_2<\frac{3}{2}$ \cite{Moon:2017rox}.     
        
        We couple the action (\ref{conformalR}) with the five dimensional conformal vector  field theory given by 
        \begin{align}
        &\hspace*{-1em}S^{(5)}_{\text{MA}}=\int d^4x dy \sqrt{-g}N\phi^{\frac{2z}{2+z}}
        \Bigg[-\frac{1}{4}{g}^{\mu\nu}
        {g}^{\rho\sigma} F_{\mu\rho} F_{\nu\sigma}-
        \frac{1 }{2}
        Z_1 N^{-2}{g}^{\mu\nu}\hat F_{5\mu}\hat F_{5\nu} \nn \\     
        &~~~-\frac{1}{2}Z_2{g}^{\mu\nu}
        \left(A_\mu+\nabla_\mu \sigma\right)\left( A_\nu+\nabla_\nu\sigma\right)
       -\frac{1}{2}Z_3 N^{-2}
        \left(\hat{A}_5+\hat{\nabla}_5 \sigma\right)\left( 
        \hat{A}_5+\hat{\nabla}_5\sigma\right)\Bigg], \label{sma}
        \end{align}
        where  ${F}_{\mu\nu}=\nabla_\mu A_\nu-\nabla_\nu A_\mu=-F_{\nu\mu}$, $\hat F_{ 5\mu}=\hat{\partial}_{5}A_{\mu}-\partial_\mu  \hat{A}_5=F_{5\mu}+N^\nu F_{\mu\nu}$, $\hat{\partial}_5=\partial_5-N^\mu \partial_\mu$,  $\hat{A}_5=A_5-N^\mu A_\mu$, $\hat{\nabla}_5 \sigma=\hat{\partial}_5\sigma$.
        For anisotropic conformal symmetry, the $Z$'s are given by 
        \begin{eqnarray}
        ~~Z_1=\phi^{4\frac{1-z}{2+z}}, ~~Z_2=\phi^{\frac{4}{2+z}}, ~~Z_3=\phi^{4\frac{2-z}{2+z}}.
        \label{relation}
        \end{eqnarray}
 Note that there exists a simple relation among $Z_1, Z_2$, and $Z_3$; $Z_1Z_2=Z_3.$         The above action (\ref{sma}) respects the following three symmetries: (i) foliation preserving diffeomorphism; along with (\ref{FPD})-(\ref{trans3}),
        \begin{align}
       & A^{'}_{\mu}(x',y')=\left(\frac{\partial x^\rho}{\partial x^{'\mu}}\right)A_{\rho}(x,y),\nn\\ &A^{'}_{5}(x',y')=\left(\frac{\partial y}{\partial y^{'}}\right)A_{5}(x,y)
+\left(\frac{\partial x^\mu}{\partial y^{'}}\right)A_\mu, ~~\sigma^{'}(x',y')=\sigma(x,y).
        \label{fpdd}\end{align}
        Under (\ref{fpdd}), one can check that
        \begin{eqnarray}
        \hat{A}^\prime _5 (x',y')=\left(\frac{\partial y}{\partial y^{'}}\right)\hat{A}_{5}(x,y),
         ~~\hat F_{5\mu}^\prime (x',y')=\left(\frac{\partial x^\nu}{\partial x^{'\mu}}\right)\left(\frac{\partial y}{\partial y^{'}}\right)\hat F_{5\nu}(x,y).
        \end{eqnarray}
        (ii) anisotropic conformal symmetry: along with (\ref{confotrans11}), with $A_\mu$, $A_5$, and $\sigma$ unaffected.
       % \begin{eqnarray}
       % A_{\mu}\rightarrow e^{-\frac{z}{2}\omega}A_\mu, ~~A_{5}\rightarrow %e^{-\frac{z}{2}\omega}A_5, ~~\sigma\rightarrow e^{-\frac{z}{2}\omega}\sigma.
 %       \end{eqnarray}
        
        \noindent (iii) gauge symmetry:
        \begin{eqnarray}
        A_\mu\rightarrow A_\mu+\nabla_\mu\Lambda, ~~A_5\rightarrow A_5+\nabla_5\Lambda, ~~\sigma\rightarrow \sigma-\Lambda.
        \end{eqnarray}

        In the limit where $\phi=\phi_0=1$, all $Z$'s are equal to 1. In this case the above action (\ref{sma}) reduces to five dimensional Stueckelberg theory given by \cite{Stueckelberg:1938zz} 
        \begin{equation}
        S^{(5)}_{\text{S}}=\int d^5x \sqrt{g^{(5)}}\left[-\frac{1}{4}G^{MP}G^{NQ}F_{MN}F_{PQ}-\frac{1}{2}G^{MN}\left(A_M+\nabla_M \sigma\right)\left(A_N+\nabla_N\sigma \right)\right],
        \end{equation}
                with unit mass for the vector field. $G^{MN}$ is the inverse of $G_{MN}$ given in equation (\ref{model}).       The dependence on the parameter $z$ disappears and it respects the five-dimensional diffeomorphism invariance.    
       
        The coupled action $S^{(5)}_{CG}+S^{(5)}_{MA}$ does not contain any dimensional parameter and also exhibits  the following scaling symmetry:
        \begin{eqnarray}
       && x\rightarrow b^{-1}x, ~~y\rightarrow b^{-z}y,
        ~~\phi\rightarrow 
        b^{\frac{2+z}{2}}\phi,    N^\mu
        \rightarrow b^{z-1}N^\mu,\nonumber\\
        &&A_\mu\rightarrow b A_\mu,~~A_5\rightarrow b^zA_5,~~\sigma\rightarrow \sigma,
        \end{eqnarray}
        and the engineering dimensions are given as 
        \begin{equation}
        [x]=-1, ~~[y]=-z, ~~[\phi]=\frac{2+z}{2}, ~~[N^\mu]=z-1, ~~[A_\mu]=1,~~[A_5]=z,~~[\sigma]=0.
        \end{equation}
        We note that some of the quantities must carry  $z$-dependent scaling dimensions in order to respect the symmetries.           The anisotropic conformal invariance also determines the  conformal weights of each field given by
        \begin{equation}
        [g_{\mu\nu}]_c=2, ~~[N]_c=z, ~~[\phi]_c=-\frac{2+z}{2}, 
        ~~[N^\mu]_c=0, ~~[A_\mu]_c=[A_5]_c=[\sigma]_c=0.\end{equation}
        We assume that  a  scale appears as a consequence of spontaneous conformal symmetry breaking of the  vacuum solution, and the canonical dimensions of the fields and spacetime are recovered.

        \section{Vacuum Solution}
        
        In this section, we search for the solutions of equation of motion derived from $S^{(5)}_{CG}+S^{(5)}_{MA}$. 
        First we introduce a scale $M_*$ which sets the scale for the conformal  symmetry breaking of the vacuum solution.
        We also assume that  $y$, $\phi$, $A_\mu$, $A_5$, and $\sigma$ recover their canonical dimensions with the vacuum solution, and therefore  re-scale them via
        \begin{equation}
        y\rightarrow M_*^{-z+1}y, ~~\phi\rightarrow M_*^{\frac{2+z}{2}}\phi,
        ~~A_\mu \rightarrow M_*^{-\frac{1}{2}}A_\mu, ~~A_5\rightarrow M^{z-\frac{3}{2}}_*A_5,~~\sigma\rightarrow M^{-{\frac{3}{2}}}_*\sigma. \label{anomalous}
        \end{equation}
        after which $\phi$ becomes  a dimensionless field and $A_\mu$, $A_5$ and $\sigma$ fields carry canonical dimension $+3/2$ each.
        We start from an simple ansatz which solves them, instead of writing down the  tedious looking equations\footnote{See Ref. \cite{Kouwn:2017qet} for details.}.  For our purpose, the detailed equations of motion
        are not necessary.  Consider the following ansatz
        \begin{equation}
        g_{\mu\nu}(x,y)=\eta_{\mu\nu}, ~N=1, ~\phi=\phi_0,~ N^\mu=0,
        \label{ansatz}
        \end{equation}
        which yields $B_{\mu\nu}=C_\mu=0$ in (\ref{conformalR}) and solves the equations of motion. 

We consider fluctuations of (\ref{sma}) around this vacuum, and rescaling via (\ref{anomalous}) and focusing only on the vector part, we obtain (with $A_5\equiv \eta$)          
        %unless $z=1$ so that $v(z)=1+D/2$, i.e., the isotropic case.
        % From here on, we
        % choose $v=(2+Dz)/2$, because it simply yields the conventional 4D
        %non-minimal coupling term of $\phi^2 R^{(4)$. The case $Dz=-2$ is given %a separate treatment.}
                \begin{align}
        S=\int d^4x dy \Bigg[&-\frac{1}{4}F^2-\frac{1}{2}Z_1(0) F_{5\mu}F^{5\mu}
        -\frac{1}{2}Z_2(0)M^2_*\left(A_\mu+\frac{1}{M_*}\nabla_\mu\sigma\right)\left(A^\mu+\frac{1}{M_*}\nabla^\mu\sigma\right) \nn \\
        &~~-\frac{1}{2}Z_3(0)M^2_*\left(\eta+\frac{1}{M_*}\nabla_5\sigma\right)\left(\eta+\frac{1}{M_*}\nabla_5\sigma\right)\Bigg], \label{seff}
        \end{align}
      where the constant $\phi_0^{2z \over z+2}$ has been reabsorbed into $A_\mu, \eta,$ and $\sigma$. And $Z_i(0)\equiv Z_i(\phi=\phi_0)~ (i=1,2,3)$ of (\ref{relation}).  Using the gauge invariance of the above action (\ref{seff}), we add a  gauge fixing term of the form
        
        \begin{eqnarray}
        S_{\text{GF}}= \int d^4x dy\left[-\frac{1}{2}\left(\xi_1 \partial_\mu A^\mu +\xi_2 Z_1(0) \partial_5 \eta+\xi_3 Z_2(0) M_*\sigma\right)^2\right].
        \end{eqnarray}
       In order to cancel $A_\mu$-$\sigma$ and $\eta$-$\sigma$ mixing terms in (\ref{seff}) which are first-order in derivative, we choose  $\xi_1= \xi_2=\xi_3^{-1}\equiv\xi$.
        Adding to the action (\ref{seff}) yields the following effective action        
        \begin{eqnarray}
        S_{eff}=M_* \int d^4x dy \left[-\frac{1}{2}A_MK^{MN}A_N+\frac{1}{2}\sigma L \sigma\right],
        \end{eqnarray}
        where   
        \begin{eqnarray}
                K^{\mu\nu}&=&-\left[ \left(\Box+Z_1(0)\partial^2_5-Z_2(0)M^2_*\right)\eta^{\mu\nu}-\left(1-\xi^2\right)\partial^\mu \partial^\nu\right]\nonumber\\
K^{\mu 5}&=&K^{5\mu}=Z_1(0)\left(1-\xi^2\right)\partial^5 \partial^\mu\nonumber\\
K^{55}&=&- Z_1(0)\left[ \Box+\xi^2 Z_1(0)\partial^2_5-Z_2(0)M^2_*\right]\label{inversemtx}
        \end{eqnarray}
        and $L=Z_2(0)(\Box+Z_1(0)\partial^2_5-\xi^{-2} Z_2(0) M^2_*)$.
        %Here we rescaled $A_5\rightarrow \sqrt{Z_1}A_5$, $\sigma\rightarrow
        %\sqrt{Z_2}\sigma$ to resume the canonical kinetic term. 
        Here, we used the relation $Z_1(0)Z_2(0)=Z_3(0).$ 

Calculation of the inverse matrix $K^{MN}$ gives the following propagator in momentum space (with $\bar k_5\equiv \sqrt{Z_1(0)}k_5, ~\bar M_*\equiv{\sqrt Z_2(0)}M_*$),
        \begin{align}
        G_{\mu\nu}&=\frac{1}{k^{2}+\bar k_5^2+\bar M_*^2}\Bigg[\eta_{\mu\nu}-\frac{\xi^2-1}{\left(\xi^2 k^2 +\xi^2 \bar k_5^2+\bar M_*^2 \right)}k_\mu k_\nu\Bigg], \label{26} \\
        G_{\mu\eta}&=\frac{(1-\xi^2)}{\left(\xi^2 k^2+\xi^2 \bar k_5^2+\bar M_*^2 \right)\left(k^2+\bar k_5^2+\bar M_*^2 \right)}k_\mu \bar k_5,  \label{27}\\
        G_{\eta\eta}&=\frac{\xi^2 k^2+\bar k_5^2+\bar M_*^2}{\left(\xi^2 k^2+\xi^2 \bar k_5^2+\bar M_*^2 \right)\left(k^2+\bar k_5^2+\bar M_*^2 \right)},\label{28}\\
G_{\sigma\sigma}&=-\frac{1}{(k^2+\bar k_5^2+\xi^{-2} \bar M_*^2)}.
        \end{align}
    Here, we rescaled $\eta\rightarrow \eta/{\sqrt{Z_1(0)}}$ and $\sigma \rightarrow \sigma/\sqrt{Z_2(0)}$    to recover the canonical form for the kinetic energy. If we take $\xi=0$, we find that Eqs. (\ref{26})-(\ref{28}) combine into (with $p_M=(k_\mu, \bar k_5)$)
\begin{eqnarray}
G_{MN}=\frac{1}{p^2+\bar M_*^2}\Bigg[\eta_{MN}+\frac{p_Mp_N}{\bar M_*^2}\Bigg],
\end{eqnarray}
and it describes the five dimensional Proca field with mass $\bar M_*^2.$ $G_{\sigma\sigma}$ decouples and this is the unitary gauge.  
  Note however that the conjugate momentum of the extra coordinate  $y$ is $ k_5$ and the dispersion relation  $p^2=k^2+Z_{1}(0)k_5^2$ suggests that the motion along fifth-direction is suppressed  by a factor of $\sqrt{Z_1(0)}.$ An extremely small value of $Z_1(0)$ could effectively localize the motion to four dimensional spacetime. For example, with $\phi_0=10$ and $z=-2.1$, $Z_{1}(0)$ of Eq. (\ref{relation}) becomes as small as $10^{-124}$. In this case, the mass of the vector field is $\bar M_*^2=10^{-40}M_*^2$. Recently, the value of $z$ has been tested in comparison with cosmological data and a
range of value of the parameter $z$ which can address the current dark energy density compared to the Planck energy density was given \cite{Kouwn:2017qet}. The upshot is that a tiny number can be generated with anisotropic factor $z$ which does not require the extreme fine-tuning. The propagators become particularly transparent in the Feynman gauge with $\xi=1$. In this case, $G_{\mu\eta}=0$ and the gauge field $A_\mu$ decouples from the scalar field $\eta$. 
All fields propagate with the same mass $M_*^2$ and $\eta$ supplies longitudinal polarization to $A_\mu$ field. The choice $\xi=\infty$ gives a transverse propagator in the sense that $p^MG_{MN}=0$.
The mass term of the $\sigma$ field disappears.

%$G_{\mu\nu}$ develops a pole at $k^2+Z_1 k_5^2=-Z_2 M^2_*$. 
%Then $Z_1=1$, this is the usual propagation of massive vector %field. However when $Z_1$ is very small, the propagation along %the extra dimension is suppressed. 

        \section{Conclusion and Discussions}
        In this paper, we constructed five dimensional massive vector-gravity theory with an anisotropic conformal invariance and demonstrated the possibility of obtaining a vector particle with a very slight mass compared with the physical mass scale which emerges as a consequence of spontaneous conformal symmetry breaking. At the same time, the built-in anisotropy can suppress highly the motion along the extra dimension effectively localizing it to four spacetime. 
        
        This approach can be contrasted to the mainstream research on the theory of the higher spacetime dimensions which treats the four dimensional spacetime and the extra dimensions on equal footing. The isotropic spacetime is more appealing from the aesthetical point of view, and we note that the anisotropic approach can be reconciled if we consider the flow of the anisotropic factor $z.$ For example, the theory starts from $z\neq 1$ at high energy but it relaxes to $z=1$ at low energy with a full spacetime symmetry. These aspects were considered in  critical behavior of gravity \cite{horava, zee}.
% but this has never been experimentally verified. We note, however, that %the two approaches can be reconciled if we considerflow f the anisotropic %factor $z.$
        
        The five dimensional operator $\Box+Z_1\partial^2_5-Z_2M^2_*$ with a $Z_1(<1)$ factor suppresses propagation along fifth dimension. For example, the static Coulomb potential with $M_*=0$ gives
        \begin{eqnarray}
        V(r,y)\sim \int dk_5 e^{ik_5y}\int d^3k\frac{e^{i\vec k\cdot \vec r}}{{\vec k}^2+Z_1k_5^2}\sim \frac{\sqrt{Z_1}r}{r(Z_1r^2+y^2)}.\label{extra}
        \end{eqnarray} 
        In the isotropic case with $Z_1=1$, the above expression reduces to four space-dimensional Coulomb potential. On the other hand, in the limit $Z_1\rightarrow 0$, it becomes 
        \begin{eqnarray}
        V(r,y)\rightarrow \frac{1}{r}\delta(y),
        \end{eqnarray}         
        completely obviating the possible motion along the fifth dimension. Therefore, the last expression in (\ref{extra}) naturally interpolates between the four and five dimensions. It gives an interesting possibility that for a very small  value of  $Z_1$, the effective motion can be practically considered to be four-dimensional. This can render the extra dimension almost completely obsolete and making its presence fall into desuetude.

 In effective four dimensional theory, the size $l_E$ of the extra dimension can be shown to satisfy the following relation from (\ref{conformalR}) after the conformal symmetry breaking
\begin{eqnarray}
l_E=l_*\phi_0^{-2}\Bigg(\frac{M_{pl}}{M_*}\Bigg)^2,
\end{eqnarray} 
where $l_*$ is the size associated with the symmetry breaking scale $M_*$ and $M_{pl}$ is the four dimensional Planck scale. When $M_*$ is close to the Planck scale, $l_E \sim l_*\sim l_{pl}$ for $\phi_0 \sim {\cal {O}}(1).$ For a value of $\phi_0\sim 10^{-7}$, $M_*\sim 10^3 {\rm GeV}$ yields $l_E$ to be an order of millimeter.  A possible connection with the localization in the brane physics \cite{Maartens:2010ar} deserves further investigations.

 We conclude with a remark on the possible application to the physics of gravitational wave \cite{gravitation}. Recently, it was investigated to check whether the gravitational wave is leaking into extra  dimension and it was found that the analysis negates the existence of extra dimension \cite{ligo, holz}\footnote{See also Ref. \cite{Visinelli:2017bny} in which the question of the observed gravitational wave leaking into the extra dimension was investigated in the brane-world model.}. In the anisotropic approach, the gravitational wave equation is governed
by anisotropic five- dimensional d'Alembertian where the isotropic  operator is replaced with $\Box+Z_g\partial^2_5$, where $Z_g$ can be given as $Z_g(\phi_0)
\propto \phi_0^{-\frac{2(z-4)}{z+2}}$ in Eq. (\ref{conformalR}). This could again suppress the gravitational wave along the extra dimension and the effect is the same as to keep the wave  from leaking into the extra dimension. Note that $Z_1(\phi_0), Z_2(\phi_0),$ and $Z_g(\phi_0)$ all can produce a very small number near at $z<-2$ and $\phi_0\sim 10$, or at $z>-2$ and $\phi_0\sim 0.1$ It would be interesting if detailed comparisons with observation can be performed.

~~\\~~\\
\noindent {\bf Acknowledgement}

We like to thank Ki-Young Choi and Seokcheon Lee for useful comments. Taegyu Kim was supported by the National Research Foundation of Korea (NRF) grant funded by the Korea government (MEST)  (NRF-2018R1D1A1B07051127, NRF-2019R1A6A1A10073079)

        %%%%%%%%%%%%%%%%%%%%%%%%%%%%%%%%%%%%%%%%%%%%%%%%%%%%%%%%%%%%%%%%%%%
        

\begin{thebibliography}{99}
                        %\cite{Tu:2005ge}
                        \bibitem{Tu:2005ge}
                        L.~C.~Tu, J.~Luo and G.~T.~Gillies,
                        ``The mass of the photon,''
                        Rept.\ Prog.\ Phys.\  {\bf 68}, 77 (2005);
                        %%CITATION = RPPHA,68,77;%%
                        %39 citations counted in INSPIRE as of 14 Nov 2015
                        %\cite{Okun:2006pn}
                        %\bibitem{Okun:2006pn}
                        L.~B.~Okun,
                        ``Photon: History, mass, charge,''
                        Acta Phys.\ Polon.\ B {\bf 37}, 565 (2006)
                        [hep-ph/0602036];
                        %%CITATION = HEP-PH/0602036;%%
                        %5 citations counted in INSPIRE as of 14 Nov 2015
                        %\cite{Goldhaber:2008xy}
                        %\bibitem{Goldhaber:2008xy}
                        A.~S.~Goldhaber and M.~M.~Nieto,
                        ``Photon and Graviton Mass Limits,''
                        Rev.\ Mod.\ Phys.\  {\bf 82}, 939 (2010)
                        [arXiv:0809.1003 [hep-ph]].
                        %%CITATION = ARXIV:0809.1003;%%
                        %115 citations counted in INSPIRE as of 14 Nov 2015
                
                        %\cite{Ruegg:2003ps}
                \bibitem{Ruegg:2003ps} 
                H.~Ruegg and M.~Ruiz-Altaba,
                %``The Stueckelberg field,''
                Int.\ J.\ Mod.\ Phys.\ A {\bf 19}, 3265 (2004)
                doi:10.1142/S0217751X04019755
                [hep-th/0304245].
                %%CITATION = doi:10.1142/S0217751X04019755;%%
                %276 citations counted in INSPIRE as of 20 Feb 2020
                
                %\cite{Accioly:2010zzb}
                \bibitem{Accioly:2010zzb} 
                A.~Accioly, J.~Helayel-Neto and E.~Scatena,
                %``Combining general relativity, massive QED and very long baseline interferometry to gravitationally constrain the photon mass,''
                Phys.\ Lett.\ A {\bf 374}, 3806 (2010).
                doi:10.1016/j.physleta.2010.07.050
                %%CITATION = doi:10.1016/j.physleta.2010.07.050;%%
                %5 citations counted in INSPIRE as of 20 Feb 2020
                
                %\cite{Kouwn:2015cdw}
                \bibitem{Kouwn:2015cdw} 
                S.~Kouwn, P.~Oh and C.~G.~Park,
                %``Massive Photon and Dark Energy,''
                Phys.\ Rev.\ D {\bf 93}, no. 8, 083012 (2016)
                doi:10.1103/PhysRevD.93.083012
                [arXiv:1512.00541 [astro-ph.CO]].
                %%CITATION = doi:10.1103/PhysRevD.93.083012;%%
                %23 citations counted in INSPIRE as of 20 Feb 2020
                
                 %\cite{Moon:2017rox}
                \bibitem{Moon:2017rox}
                T.~Moon and P.~Oh,
                %``$z-$Weyl gravity in higher dimensions,''
                JCAP {\bf 1707}, no. 07, 024 (2017)
                doi:10.1088/1475-7516/2017/09/024
                [arXiv:1705.00866 [hep-th]].
                
                %\cite{Kouwn:2017qet}
                \bibitem{Kouwn:2017qet} 
                S.~Kouwn, P.~Oh and C.~G.~Park,
                %``The Effect of Anisotropic Extra Dimension in Cosmology,''
                Phys.\ Dark Univ.\  {\bf 22}, 27 (2018)
                doi:10.1016/j.dark.2018.08.003
                [arXiv:1709.08499 [astro-ph.CO]].
                %%CITATION = doi:10.1016/j.dark.2018.08.003;%%
                %2 citations counted in INSPIRE as of 20 Feb 2020
                
                %\cite{Stueckelberg:1938zz}
                \bibitem{Stueckelberg:1938zz} 
                E.~C.~G.~Stueckelberg,
                %``Interaction forces in electrodynamics and in the field theory of nuclear forces,''
                Helv.\ Phys.\ Acta {\bf 11}, 299 (1938).
                %%CITATION = HPACA,11,299;%%
                %220 citations counted in INSPIRE as of 20 Feb 2020
                
                %\cite{Horava:2009uw}
                \bibitem{horava} 
                P.~Horava, %``Quantum Gravity at a Lifshitz Point,''
                Phys.\ Rev.\ D {\bf 79}, 084008 (2009)
                doi:10.1103/PhysRevD.79.084008
                [arXiv:0901.3775 [hep-th]].
                %%CITATION = doi:10.1103/PhysRevD.79.084008;%%
                %1859 citations counted in INSPIRE as of 28 Feb 2020
                
                %\cite{Porto:2009xj}
                \bibitem{zee} 
                R.~A.~Porto and A.~Zee,
                %``Relaxing the cosmological constant in the extreme ultra-infrared,''
                Class.\ Quant.\ Grav.\  {\bf 27}, 065006 (2010)
                doi:10.1088/0264-9381/27/6/065006
                [arXiv:0910.3716 [hep-th]].
                %%CITATION = doi:10.1088/0264-9381/27/6/065006;%%
                %9 citations counted in INSPIRE as of 28 Feb 2020
                
              
                
                %\cite{Maartens:2010ar}
                \bibitem{Maartens:2010ar} 
                R.~Maartens and K.~Koyama,
                %``Brane-World Gravity,''
                Living Rev.\ Rel.\  {\bf 13}, 5 (2010)
                doi:10.12942/lrr-2010-5
                [arXiv:1004.3962 [hep-th]].
                %%CITATION = doi:10.12942/lrr-2010-5;%%
                %403 citations counted in INSPIRE as of 20 Feb 2020
               
               
               %\cite{Abbott:2016blz}
               \bibitem{gravitation} 
               B.~P.~Abbott {\it et al.} [LIGO Scientific and Virgo Collaborations],
               %``Observation of Gravitational Waves from a Binary Black Hole Merger,''
               Phys.\ Rev.\ Lett.\  {\bf 116}, no. 6, 061102 (2016)
               doi:10.1103/PhysRevLett.116.061102
               [arXiv:1602.03837 [gr-qc]].
               %%CITATION = doi:10.1103/PhysRevLett.116.061102;%%
               %4797 citations counted in INSPIRE as of 28 Feb 2020
               

%\cite{Abbott:2018lct}
\bibitem{ligo} 
B.~P.~Abbott {\it et al.} [LIGO Scientific and Virgo Collaborations],
%``Tests of General Relativity with GW170817,''
Phys.\ Rev.\ Lett.\  {\bf 123}, no. 1, 011102 (2019)
doi:10.1103/PhysRevLett.123.011102
[arXiv:1811.00364 [gr-qc]].
%%CITATION = doi:10.1103/PhysRevLett.123.011102;%%
%112 citations counted in INSPIRE as of 28 Feb 2020


%\cite{Pardo:2018ipy}
\bibitem{holz} 
K.~Pardo, M.~Fishbach, D.~E.~Holz and D.~N.~Spergel,
%``Limits on the number of spacetime dimensions from GW170817,''
JCAP {\bf 1807}, 048 (2018)
doi:10.1088/1475-7516/2018/07/048
[arXiv:1801.08160 [gr-qc]].
%%CITATION = doi:10.1088/1475-7516/2018/07/048;%%
%47 citations counted in INSPIRE as of 28 Feb 2020                
                         
%\cite{Visinelli:2017bny}
\bibitem{Visinelli:2017bny}
L.~Visinelli, N.~Bolis and S.~Vagnozzi,
%``Brane-world extra dimensions in light of GW170817,''
Phys. Rev. D \textbf{97} (2018) no.6, 064039
doi:10.1103/PhysRevD.97.064039
[arXiv:1711.06628 [gr-qc]]; S.~Vagnozzi and L.~Visinelli,
    %``Hunting for extra dimensions in the shadow of M87*,''
    Phys. Rev. D \textbf{100} (2019) no.2, 024020
    doi:10.1103/PhysRevD.100.024020
    [arXiv:1905.12421 [gr-qc]].
    %52 citations counted in INSPIRE as of 21 Jul 2020   
                
                
                
        \end{thebibliography}
\end{document}